\documentstyle[sprocl,epsf]{article}

\def\Journal#1#2#3#4{{#1}{\bf #2}, #3 (#4)}

\def\NPB{{\em Nucl. Phys.} B}
\def\PLB{{\em Phys. Lett.} B}

\def\PRD{{\em Phys. Rev.} D}
\def\SJNP{\em Sov. J. Nucl. Phys. }
\def\RMP{\em Rev. Mod. Phys. }

\def\be{\begin{equation}}
\def\ee{\end{equation}}
\def\bea{\begin{eqnarray}}
\def\eea{\end{eqnarray}}

\begin{document}

\title{NONEQUILIBRIUM NEUTRINO OSCILLATIONS AND PRIMORDIAL 
NUCLEOSYNTHESIS}

\author{ D.P. KIRILOVA~\footnote{Institute of Astronomy,
 Bulgarian Academy of Sciences,
bul. Tsarigradsko Shosse 72, Sofia, Bulgaria, 
E-mail: $mih@phys.uni$-$sofia.bg$}
}

\address{Theoretical Astrophysics Center, Juliane Maries Vej 30,
           DK-2100 Copenhagen O, Denmark\\
}

\author{ M.V. CHIZHOV }

\address{Centre for Space Research and Technologies, Physics Department,\\
Sofia University, bul. James Bourchier 5, Sofia, Bulgaria\\
E-mail: $mih@phys.uni$-$sofia.bg$}

\maketitle
\abstracts{
We studied nonequilibrium oscillations between  
left-handed electron neutrinos and nonthermalized sterile neutrinos.
The exact kinetic equations for neutrinos,
written in terms of neutrino density matrix
in {\it momentum} space were analyzed.
The evolution of neutrino density matrix was 
numerically calculated. This allowed  
to study precisely the evolution of the neutrino number densities, 
energy spectrum distortion and the asymmetry between neutrinos and 
antineutrinos for each momentum mode. Both effects of distortion and 
asymmetry,
which cannot be accounted for correctly when working in terms of 
particle densities and mean energies, and 
the depletion of 
electron neutrino state 
have been proved considerable 
for a certain range of oscillation parameters. 
The influence of nonequilibrium oscillations on primordial nucleosynthesis
 was calculated. Cosmologically excluded regions for 
oscillation parameters were obtained.}

\section{Nonequilibrium neutrino oscillations}
 
   We discuss nonequilibrium oscillations between weak interacting 
electron neutrinos $\nu_e$ and sterile neutrinos $\nu_s$ 
for the case when $\nu_s$ do not thermalize till 2 $MeV$
and oscillations become effective after $\nu_e$ decoupling. 
Oscillations of  that type, but for the case of $\nu_s$ thermalizing 
before or around 2 $MeV$ have been already discussed in 
literature.\cite{bd1}$^{\!-\,}$\cite{sv} We have provided a proper kinetic 
analysis of the neutrino evolution in terms of {\it kinetic equations
  for the neutrino density matrix in momentum space}. 
The assumptions of the model are the following: (a) 
Singlet neutrinos decouple much earlier than 
the active neutrinos do:  $T^F_{\nu_s} \ge T^F_{\nu_e}$,
therefore, in later epochs  $T_{\nu_s} \le T_{\nu_e}$, 
 due to the additional heating of $\nu_e$
in equilibrium in comparison with the already decoupled $\nu_s$. 
Hence, the number densities of  $\nu_s$ are considerably less 
than those of $\nu_e$, $N_{\nu_s}<<N_{\nu_e}$. 
(b)We consider oscillations between $\nu_s$ and $\nu_e$, according to 
the Majorana\&Dirac mixing scheme~\cite{b} 
with mixing present just in the electron sector
$\nu_i={\cal U}_{il}\nu_l$, $l=e,s$:\footnote{The transitions 
between different 
neutrino flavours were proved to have negligible effects.\cite{do}}
\be
\begin{array}{ccc}
\nu_1 & = &  cos(\vartheta)\nu_e+sin(\vartheta)\nu_s\\
\nu_2 & = & -sin(\vartheta)\nu_e+ cos(\vartheta)\nu_s,
\end{array}
\ee
where $\nu_s$ denotes the sterile electron antineutrino, $\nu_1$ and 
 $\nu_2$ are Majorana particles with masses correspondingly $m_1$ and $m_2$. 
(c)We assume that neutrino oscillations become 
effective after the decoupling 
of the active neutrinos, $\Gamma_{osc}\ge H$ for $T\le 2 MeV$. 
This puts constraint on the neutrino mass difference:
 $\delta m^2 \le 1.3 \times 10^{-7}$ $eV^2$. 
(d) We require that sterile neutrinos have not thermalized till 
2 $MeV$ when oscillations become effective. 
 This puts the following limit on the allowed 
range of oscillation parameters:
\cite{ma}$^{\!,\,}$\cite{bd1}$^{\!-\,}$\cite{ekt}
 $sin(2\vartheta) \delta m^2 \le 10^{-7} eV^2$.

 As far as for this model 
the rates of expansion of the Universe, neutrino oscillations and 
neutrino interactions with the medium may be comparable, we  
have used kinetic equations for neutrinos accounting {\it simultaneously} 
for the participation of neutrinos into expansion, oscillations and 
interactions with the 
medium.\cite{do}$^{\!,\,}$\cite{dpk}$^{\!,\,}$\cite{sr}
We have analyzed the evolution of nonequilibrium oscillating 
neutrinos by numerically integrating the kinetic equations for the 
density matrix in {\it momentum} space for the period after the decoupling 
of the electron neutrino till the freezing of neutron-proton ratio
($n/p$-ratio), i.e. for $2~ MeV \ge T \ge 0.3~ MeV$.  
 We considered both resonant $\delta m^2 = m_2^2 - m_1^2 <0$ 
 and nonresonant $\delta m^2 >0$ oscillations.

 \section{ Kinetics of nonequilibrium neutrino oscillations}
 
The kinetic equations for the density matrix of the nonequilibrium 
oscillating neutrinos in the primeval plasma of the Universe 
in the epoch previous to nucleosynthesis have the form:
\be
{\partial \rho(t) \over \partial t} =
H p~ {\partial \rho(t) \over \partial p} 
+ i \left[ {\cal H}_o, \rho(t) \right]
+i \left[ {\cal H}_{int}, \rho(t) \right] 
+ {\rm O}\left({\cal H}^2_{int} \right),
\label{kin}
\ee
where $p$ is the momentum of neutrino
and $\rho$ is the density matrix of the massive Majorana 
neutrinos in momentum space.

The first term in the right side of Eq.\ref{kin}
describes the effect of expansion, 
the second is responsible for oscillations, the 
third accounts for forward neutrino scattering off the 
medium.\cite{la}
${\cal H}_o$ is the free neutrino Hamiltonian:
\be
{\cal H}_o = \left( \begin{array}{cc}
\sqrt{p^2+m_1^2} & 0 \\ 0 & \sqrt{p^2+m_2^2}
\end{array} \right),
\ee
while ${\cal H}_{int} = \alpha~V$ is the interaction Hamiltonian, 
where $\alpha_{ij}=U^*_{ie} U_{je}$, 
$V=G_F \left(\pm L - Q/M_W^2 \right)$,
and in the interaction basis has the form 
\be
{\cal H}_{int}^{LR} = \left( \begin{array}{cc}
V & 0 \\ 0 & 0 \end{array} \right).
\ee
The first `local' term in $V$ accounts 
for charged- and neutral-current
tree-level interactions with medium protons, neutrons,
electrons and positrons, neutrinos and antineutrinos. 
It is proportional to  
the fermion asymmetry of the plasma $L=\sum_f L_f$, which is assumed 
of the order of the baryon one  
\be
L_f \sim {N_f-N_{\bar{f}} \over N_\gamma}~T^3 \sim
{N_B-N_{\bar{B}} \over N_\gamma}~T^3 = \beta T^3.
\ee
 The second `nonlocal' term arises as an $W/Z$ 
 propagator effect,\cite{nr}
$Q \sim E_\nu~T^4$.
The two terms have different 
temperature dependence and an interesting interplay between them 
during the cooling of the Universe is observed. 
The last term in the Eq.\ref{kin} describes the weak interactions 
of neutrinos with the 
medium.\footnote{ For example, for the weak reactions 
of neutrinos with electrons and positrons $e^+ e^- \leftrightarrow
\nu_i \tilde{\nu}_j$, $e^\pm \nu_j \to e'^\pm \nu'_i$
it was explicitly written in Refs.~10,12.} 

We have analyzed the evolution of the neutrino density matrix 
assumed that oscillations become noticeable 
after electron neutrinos decoupling. So, the neutrino kinetics 
down to 2 $MeV$ does not differ from the standard case, i.e. 
electron neutrinos maintain their equilibrium distribution, 
while sterile neutrinos are absent. 
Then the last term in the kinetic equation can 
be neglected.
The equation results into a set of coupled nonlinear 
integro-differential equations for the components of the density 
matrix.\footnote{For the case of vacuum neutrino oscillations 
these equations were analytically solved in Ref.~12.}
 We have numerically calculated the evolution of the 
neutrino density matrix for the temperature interval 
$[0.3,2.0]~ MeV$. The oscillation parameters range 
studied is $\delta m^2 \in \pm [10^{-10}, 10^{-7}]~eV^2$
 and $\vartheta \in[0,\pi/4]$. 
The baryon asymmetry $\beta$ was taken to be $3\times 10^{-10}$.

\section{Nucleosynthesis with nonequilibrium oscillating neutrinos}

We analyzed the influence of nonequilibrium oscillations on the 
primordial production of $^4\! He$. The effect of oscillations on 
nucleosynthesis has been discussed in 
numerous 
publications.\cite{bd1}$^{\!-\,}$\cite{sv}$^{\!,\,}
$\cite{dpk}$^{\!,\,}$\cite{hp} 
 Here we provided a detail kinetic 
calculations of helium abundance for the case of nonequilibrium 
oscillations in medium. The kinetic equation describing the evolution 
of the neutron number density in momentum space $n_n$ for the case of 
oscillating neutrinos $\nu_e \leftrightarrow \nu_s$ was numerically 
integrated for the temperature range of interest $T \in [0.3,2.0]~ MeV$.

\section{Results and conclusions}

 Our numerical analysis showed that 
the nonequilibrium oscillations 
can considerably deplete the number densities of 
electron neutrinos (antineutrinos), distort their 
energy spectrum and produce neutrino-antineutrino asymmetry  that 
may grow at the resonant transition and may change 
considerably the evolution 
of neutrino ensembles. The effects of nonequilibrium oscillations on 
nucleosynthesis may be considerable for certain range of oscillation 
parameters.
The results of our study are as follows:

(a) As far as oscillations become effective when
the number densities of $\nu_e$ 
are much greater than those of $\nu_s$, $N_{\nu_e} \gg N_{\nu_s}$,
the oscillations tend to reestablish the statistical 
equilibrium between different oscillating species. 
As a result $N_{\nu_e}$ decreases in comparison to its standard  
equilibrium value.\cite{bd2}$^{\!-\,}$\cite{s} 

The depletion of the electron neutrino number densities due to 
oscillations to sterile ones leads to an effective decrease in the 
weak processes rates, and thus to an increase of the freezing  
temperature of the $n/p$-ratio and corresponding overproduction of the 
primordially produced $^4\! He$.

\begin{figure}
\vskip -2.2truecm
\hspace{10truecm}
\epsfysize=11truecm\epsfbox[25 -180 325 210]{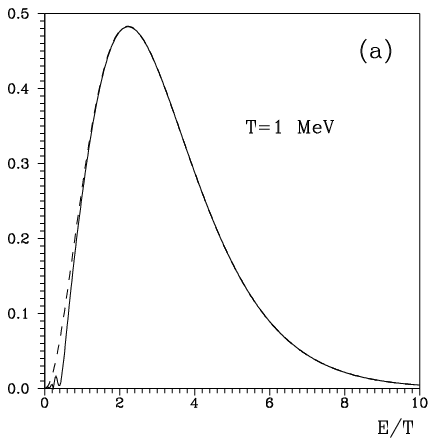}
\epsfysize=11truecm\epsfbox[170 -180 470 210]{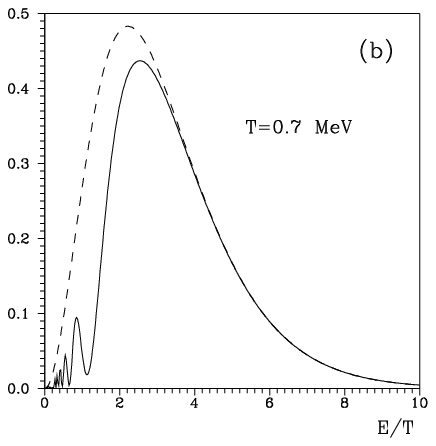}
\epsfysize=11truecm\epsfbox[315 -180 615 210]{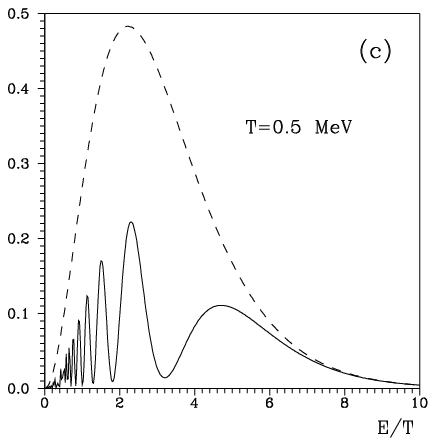}
\vspace{-5.5truecm}
\caption{The figure shows the energy distortion of active
neutrinos $x^2 \rho_{LL}(x)$, where $x=E_\nu/T$, for the case of
nonequilibrium neutrino oscillations, 
$\delta m^2 = -10^{-8}$, $\vartheta=\pi/8$
at different temperatures:
$T=1~MeV$ (a), $T=0.7~MeV$ (b), and $T=0.5~MeV$ (c).}
\label{specter}
\end{figure}

(b) For the case of strongly nonequilibrium oscillations 
the distortion of the energy distribution of neutrinos may be 
considerable.\footnote{This effect was discussed for a first 
time by Dolgov,\cite{do}
 but unfortunately, as far as the case of flavour neutrino 
oscillations were considered and the energy distortion for that case was 
shown to be negligible, it was not paid the necessary attention it 
deserved. In paper by Kirilova~\cite{dpk} 
it was first shown that for the case of 
$\nu_e \leftrightarrow \nu_s$ {\it vacuum oscillations}  this effect 
is considerable.} The evolution of the distortion is the following:
First the low energy part of the spectrum is distorted, and later on 
this distortion concerns neutrinos with higher and higher 
energies (Fig.~\ref{specter}).

 This behavior is 
natural, as far as neutrino oscillations affect first low energy neutrinos,
$\Gamma_{osc} \sim \delta m^2/E_{\nu}$. 
 The naive account of this effect by shifting the effective temperature and 
assuming the neutrino spectrum of equilibrium form gives wrong results 
for the case $\delta m^2 < 10^{-7} eV^2$. 
 
 The effect of the distortion on primordially produced helium is as 
follows. An average 
decrease of the energy of active neutrinos leads to a 
decrease of the weak reactions rate, $\Gamma_w \sim E_\nu^2$ and 
subsequently to an increase in the freezing temperature and the 
produced helium. On the other hand, there exists an energy 
threshold for the reaction $\tilde{\nu_e}+p \to n+e^+$. And in 
case when, due to oscillations, the energy of the relatively 
greater part of neutrinos becomes smaller than that threshold 
the $n/p$- freezing ratio decreases.\cite{ki}
The numerical analysis showed that the latter effect is 
less noticeable compared with the previously described ones. 
 
(c) Other interesting effect revealed by our approach is the  
generation of asymmetry between $\nu_e$ and their antiparticles. 
The possibility of an asymmetry generation
due to oscillations was discussed in many 
papers.\cite{hp}$^{\!,\,}$\cite{la}$^{\!,\,}$\cite{bd1}$^{\!-\,}
$\cite{ekm}$^{\!,\,}$\cite{ftv}$^{\!,\,}$\cite{s}
Our approach allowed precise description of the asymmetry and its 
evolution, as far as working with the {\it self-consistent 
kinetic equations 
for neutrinos in momentum space} enabled us to calculate the 
behavior of asymmetry at each momentum. The calculated 
result may differ considerably from the rough 
estimations made by working with neutrino mean energy and with the 
integrated quantities like particle densities and 
the energy densities.  
The asymmetry effect is noticeable only 
for the resonant case.  Even when the asymmetry 
is assumed initially negligibly small (of the order of the baryon one), 
i.e. $\sim 10^{-10}$, 
it may be considerably amplified at resonant transition due to different 
interactions of neutrinos and antineutrinos with the CP-odd medium. 
 The value of the asymmetry may grow 
by several orders of magnitude, oscillating and sign changing. 
 Even in case when the value of the asymmetry 
does not become considerable enough to have some direct noticeable effect, 
on primordial nucleosynthesis for example, 
the asymmetry term at the resonant transition
 determines the evolution of the neutrino density matrix. 
 It effectively suppresses the resonant transitions of active neutrinos (antineutrinos)
thus weakening neutrino depletion at resonance.  
 For some model parameters this effect 
consists 20\% of the previously discussed. 

The asymmetry calculations showed a slight predominance of neutrinos 
over antineutrinos, leading to decrease of helium. The greater effect 
of the asymmetry is, however, the change in the evolution of the neutrino 
ensembles during and after the resonance resulting to 
a relative increase of both neutrino 
and antineutrino particle densities. This may lead to a noticeable  
underproduction of helium (up to $10\%$ relative decrease). 

\begin{figure}
\vskip -0.9truecm
\hspace{10truecm}
\epsfysize=11truecm\epsfbox[-50 -275 300 225]{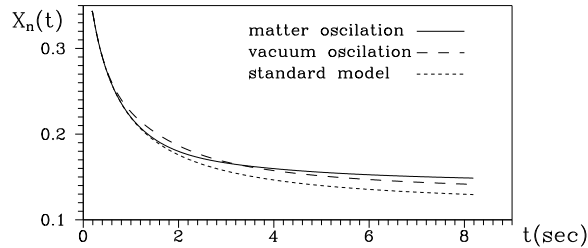}
\vspace{-7truecm}
\caption{ The curves represent the evolution of the neutron
number density relative to nucleons $X_n(t)=N_n(t)/(N_p+N_n)$ for
the nucleosynthesis model with vacuum nonequilibrium oscillations
and for the case of nonequilibrium oscillations in medium,
$\delta m^2 = -10^{-8}$, $\vartheta=\pi/8$. For comparison the curve
corresponding to the standard nucleosynthesis model is shown.}
\label{he}
\vskip -0.5truecm
\end{figure}

The total effect of nonequilibrium neutrino oscillations is 
overproduction of helium in comparison to the standard value. 
The results of the 
numerical integration are illustrated on Fig.~\ref{he}, 
in comparison with the vacuum case and the standard nucleosynthesis 
without oscillations. 

\begin{figure}[h]
\vskip -1.9truecm
\hspace{10truecm}
\epsfysize=11truecm\epsfbox[-50 -250 300 250]{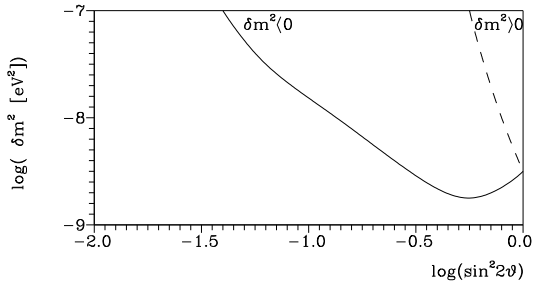}
\vspace{-6truecm}
\caption{Exclusion regions for oscillation parameters are shown 
for the case of resonant $\delta m^2 < 0$ and nonresonant $\delta m^2 > 0$
neutrino oscillations. The curves correspond to helium abundance 
$Y_p=0.245$.}
\label{excl}
\end{figure}

From numerical integration for different oscillation parameters 
we have obtained constant $^4\! He$ contours.  
We have used the $4\%$ relative increase in the primordially produced 
helium to obtain the exclusion region for the oscillation parameters 
(Fig.~\ref{excl}). 
 
For the cases when the energy 
distortion and asymmetry are considerable we have obtained an 
order of magnitude stronger constraints than the cited in 
literature.\cite{ekt}$^{\!-\,}$\cite{s}
 Therefore, in conclusion we would like to 
stress once again, 
that in case of nonequilibrium neutrino oscillations working 
with the exact kinetic equations for the density matrix of 
neutrinos in momentum space is necessary. 

\section*{Acknowledgments}

 We are glad to thank the 
Theoretical Astrophysical Center where the preprint version of 
the work was prepared for the warm hospitality and financial support. 

The authors thank Professor Matts Roos and the organizing committee 
of the NEUTRINO 96 Conference for the possibility to present 
the results of this work there.

This work was supported in part by the Danish National Research
Foundation through its establishment of the Theoretical Astrophysics Center.

\end{document}